# Structural origin of the $J_{\text{eff}}$=1/2 antiferromagnetic phase in Ga-doped $Sr_2IrO_4$

H. W. Wang[1], L. Y. Zhang[1], N. Hu[2], B. You[1], Y. T. Chang[1], S. L. Yuan[1], C. L. Lu[1*], and J. –M. Liu[3,4]

[1] *School of Physics & Wuhan National High Magnetic Field Center, Huazhong University of Science and Technology, Wuhan 430074, China*

[2] *School of Science, Hubei University of Technology, Wuhan 430068, China*

[3] *Laboratory of Solid State Microstructures, Nanjing University, Nanjing 210093, China*

[4] *Institute for Advanced Materials, Hubei Normal University, Huangshi 435001, China*

---

[*] Email: cllu@hust.edu.cn

# Abstract

$Sr_2IrO_4$ hosts a novel $J_{eff}$ =1/2 Mott state and quasi-two-dimensional antiferromagnetic order, providing a unique avenue of exploring emergent states of matter and functions that are extraordinarily sensitive to any structural variations. While the correlation between the physical property and lattice structure in $Sr_2IrO_4$ has been a focused issue in the past decade, a common perception assumes that the magnetic ordering is essentially determined by the Ir-O-Ir bond angle. Therefore, a delicate modulation of this angle and consequently a major modulation of the magnetic ordering, by chemical doping such as Ga at Ir site, has been extensively investigated and well believed. In this work, however, we present a whole package of structure and magnetism data on a series of single crystal and polycrystalline $Sr_2Ir_{1-x}Ga_xO_4$ samples, revealing the substantial difference in the Néel temperature $T_N$ between the two types of samples, and the $T_N$ value for the polycrystalline sample $x$ = 0.09 is even 64 K higher than that of the single crystal sample $x$ = 0.09 ($\Delta T_N \sim$ 64 K at $x$ = 0.09). Our systematic investigations demonstrate the crucial role of the $c/a$ ratio in tuning the interlayer coupling and thereby the Néel point $T_N$, i.e. a higher $T_N$ can be achieved as $c/a$ is reduced. The notable differences in structural parameters between the two groups of samples are probably caused by additional strain due to the massive grain boundaries in polycrystalline samples. The present work suggests an additional ingredient of physics that is essential in modulating the emergent properties in $Sr_2IrO_4$ and probably other iridates.

## I. Introduction

The family of 5$d$ iridates has been an emergent group of quantum materials for studying new phases of matter arising from the large relativistic spin-orbit coupling (SOC, $\lambda \sim 0.5$ eV) that is comparable to other fundamental interactions such as Hubbard interaction $U$ and electronic bandwidth $W$ [1-6]. A growing list of nontrivial quantum states due to a delicate interplay of the relevant energies have so far been revealed in iridates, including the $J_{eff}$=1/2 Mott state, Weyl semimetals with Fermi arcs, correlated topological insulator, Kitaev spin liquid, excitonic magnetism of pentavalent $Ir^{5+}$ ($5d^4$), etc [2]. Among these, the $J_{eff}$=1/2 state, a profound manifestation of the strong SOC, has been a rare but unique example that has been better understood both theoretically and experimentally [7].

As a prototype, $Sr_2IrO_4$ was first revealed to host the $J_{eff}$=1/2 Mott state arising from the cooperation of crystal field, SOC, and Coulomb repulsion. A signature character of $Sr_2IrO_4$ could be the close correlation between physical properties and lattice degree of freedom [8-12], sharply differing from conventional 3$d$ compounds. Theoretical calculations revealed that the ground state of $Sr_2IrO_4$ can move away or toward the $J_{eff}$=1/2 limit simply by varying some structure parameters, noting that the canonical $J_{eff}$=1/2 state can only exist in a cubic crystal environment [13]. Specifically, Moon *et al.* predicted that the electronic structure of $Sr_2IrO_4$ can be renormalized by changing the in-plane Ir-O-Ir bond angle $\phi$, i.e. a smaller band gap is expected when $\phi$ is increased [9]. Another typical example is the so-called locking effect, which describes a unique relationship between the canting angle $\alpha$ of $J_{eff}$=1/2 magnetic moment and rotation angle $\varphi$ of $IrO_6$ octahedron, i.e. $\alpha \sim \varphi$ [8,14,15]. Since $\varphi$ has the one to one correspondence with $\phi$ (i.e. $\phi+2\varphi=180°$), tilting the $J_{eff}$=1/2 moment may lead to modification of the band gap. This is confirmed by several recent studies focusing on the anisotropic magnetoresistance of $Sr_2IrO_4$ [16-18]. In shorts, it has been surprising to observe such a simple scenario for the relationship between the crystalline structure and electronic property in such a system where the complicated cooperation of multiple fundamental interactions is available.

Although this scenario has been widely utilized to interpret the properties of $Sr_2IrO_4$, it has encountered challenge to understand the long range antiferromagnetic (AFM) order. First, since both the band gap and $J_{eff}$= 1/2 moment are associated with $\phi$, it is natural to assume intimate coupling between the Mott state and AFM phase. This is in parallel with the fact that the magnetic interaction

(~ 0.1 eV) and band - gap (~ 0.3 eV) of $Sr_2IrO_4$ have similar energy scale [19,20]. However, some experimental results unambiguously show decoupling of the AFM phase and the Mott state [21,22], deviating from this assumption. Such discrepancy may mainly due to the elusive relationship between the Néel order and crystalline structure. Second, $Sr_2IrO_4$ is a single - layer perovskite, belonging to the Ruddlesden - Popper series. Despite the intralayer exchange interaction is strong and primarily determined by the in-plane Ir-O-Ir bond angle $\phi$, the 3 - dimensional AFM order is stabilized by the much weaker interlayer coupling (~ 1 µeV), resembling the situation of $La_2CuO_4$ [23]. Recent theoretical studies have confirmed this picture, and further proposed that the tetragonal distortion is more essential to the interlayer coupling [10,11,24]. While these theories have unraveled interesting aspects of the AFM phase of $Sr_2IrO_4$, the related experiments are largely unexplored.

Chemical substitution has been demonstrated to be very efficient in modulating both the Néel order and crystalline structure of $Sr_2IrO_4$ [25-29], and thus is one of the promising approaches to investigate the structural origin of the AFM phase. Our previous work revealed that Ga-substitution at the Ir-site in $Sr_2IrO_4$ can cause evident variation in both the lattice and the Néel point ($T$) $T_N$ [30], providing advantages to address this issue. Importantly, $Ga^{3+}$ is nonmagnetic, and thus doesn't introduce magnetic fluctuations into the system, simplifying the study greatly. Regarding the carrier-doping effect due to the lower valance state of $Ga^{3+}$ than $Ir^{4+}$, it can be ignored by simply making a comparison study between single crystal and polycrystalline $Sr_2Ir_{1-x}Ga_xO_4$ where the doping effect would be identical. Here, it is worth mentioning that involving polycrystalline samples may lead to further modulation of the lattice and properties, and thus facilitating the comparison. For instance, polycrystalline and single crystal $(Sr_{1-x}La_x)_2IrO_4$ were found to show different lattice parameters and properties [29,31-34].

In the present work, two groups of $Sr_2Ir_{1-x}Ga_xO_4$ samples, including both single crystals with $0 \leq x \leq 0.09$ and polycrystals with $0 \leq x \leq 0.15$, have been synthesized. Extensive characterizations such as the crystalline structure, magnetism, and electric transport have been performed on all samples. Our experiments identified dramatically different physical properties between the two groups of samples, which are closely associated with their different lattice parameters. In particular, we found that the polycrystalline samples host generally higher $T_N$ than the single crystals, and the difference between the two groups of samples can be as large as $\Delta T_N \sim 64$ K at $x=0.09$. Structural analysis revealed that the $c/a$ ratio plays a crucial role in tuning $T_N$.

## II. Experimental details

A series of $Sr_2Ir_{1-x}Ga_xO_4$ ($0 \leq x \leq 0.09$) single crystals were grown using flux method from high-purity $SrCl_2$, $SrCO_3$, $IrO_2$ and $Ga_2O_3$. Pt crucible covered with lid was used for the single crystal growth [30]. Black plate-shaped crystals with typical dimension $1.2 \times 0.5 \times 0.2$ mm$^3$ were obtained after dissolving $SrCl_2$ with distilled water. For comparison, an additional group of polycrystalline samples $Sr_2Ir_{1-x}Ga_xO_4$ ($0 \leq x \leq 0.15$) were synthesized using conventional solid-state sintering method. Stoichiometric amounts of high purity $SrCO_3$, $IrO_2$, $Ga_2O_3$ were mixed thoroughly, and then calcined at temperatures ($T$) ranging from 1050 °C to 1200 °C for 24 hours each with intermediate grinding sequences. Finally, the obtained black powder was pressed into pellets and sintered at 1250 °C for 72 hours in air.

Crystalline structure of the two groups of samples were characterized by performing powder X-ray diffraction (XRD) at room temperature. Before the XRD measurements, all samples including the single crystals and polycrystals were crashed thoroughly. Elemental mappings using the energy dispersive X-ray spectrometer (EDS) were carried out in a Sigma 500 of field emission scanning electron microscope (SEM). The valance state of Ir cation was determined using x-ray photoelectron spectroscopy (XPS). Detailed measurements on magnetization ($M$) as a function of $T$ and magnetic field $H$ were carried out using a superconducting quantum interference device (SQUID, Quantum Design). During the $T$-dependence of $M$ measurements, the measuring field was fixed at $H$=0.1 T. For the $c$-axis magnetization ($M_c$) measurements, the sample was fixed on the top of a quartz rod properly to avoid any misalignment, which was then mounted to a copper stick with a hollow core. Low-temperature glue was used to fix the sample. Temperature dependence of resistivity ($\rho$) were measured using a standard four-electrode method in a physical property measurement system (PPMS, Quantum Design).

## III. Experimental results
## A. Structural characterization

All samples are pure phase with a tetragonal symmetry, confirmed by the collected XRD spectra. To acquire more details of the crystalline structure, Rietveld-profile refinements of the XRD patterns are performed for all samples. Additional functions were involved in the refinements of single crystal samples, in order to solve the problems arising from preferred orientation. The refined result for Sr$_2$IrO$_4$ single crystal is presented in Fig. 1(a). The difference between the measured and refined spectra is small with a reliability factor $R_{wp}$=3.98%. For all other samples, the refinements are all high quality, and the $R_{wp}$ values are at similar levels. The derived lattice parameters of Sr$_2$IrO$_4$ single crystal are $a$= 5.4933 Å, $c$=25.8031 Å, and $\phi$=156.3°, which are in good agreement with previous works [25]. For polycrystalline Sr$_2$IrO$_4$, the parameters are $a$= 5.5009 Å, $c$=25.8164 Å, and $\phi$=158.1°. The lattice constants $a$ and $c$ are nearly identical for the two samples, while $\phi$ of the polycrystalline Sr$_2$IrO$_4$ is about 1.1% larger than the single crystal. Similar discrepancy has been reported in previous studies [29,35].

The derived structural parameters such as $a$, $c$, $\Delta V$, $c/a$, in-plane Ir-O bond length, $\phi$ and $\varphi$ as a function of Ga-content $x$ are plotted in Fig. 1(b)-(h), respectively. In a qualitative sense, the parameters show similar tendency with $x$ for both groups of samples. For instance, $a$ increases but $c$ decreases with $x$ successively, and $\phi(x)$ shows a maximum at $x$~0.05 for both groups of samples. The anisotropic variation of $a$ and $c$ give rise to continuous decrease in the $c/a$ ratio, shown in Fig. 1(e). However, the shrink of the $c$-axis is as striking as ~ -1.5% at $x$=0.15 for the polycrystalline samples, far larger than that of the single crystals (~-0.24% at $x$=0.09). As a consequence, remarked reductions in the $c/a$ ratio and cell volume $\Delta V$ are induced in the polycrystalline samples. In the meanwhile, much larger variation in $l$, $\phi$, and $\varphi$ are also evidenced for the polycrystalline samples as compared with the single crystals, i.e. $\Delta\phi/\phi$~3.42% for the former and $\Delta\phi/\phi$~2.24% for the later. The remarkable variations in the lattice parameters evidence the active role of Ga cation in tuning the lattice of Sr$_2$IrO$_4$ as compared with other dopants [25,36].

The rigorous structural modulation in the polycrystalline samples is intrinsic, as confirmed by additional characterizations such as EDS and XPS measurements. EDS characterizations were utilized to check elemental distribution and composition of the samples. It is evidenced that all samples are stoichiometric and homogeneous without showing any aggregation of ions. Here we take the sample with $x$=0.07 as an example, and show the SEM images and EDS mappings in Fig. 2. While the single crystal sample has a uniform and smooth surface morphology, the polycrystalline sample possesses plentiful grains. Importantly, according to the EDS mappings, all elements are homogeneously distributed in the samples, and there is no contrast between grains and grain boundaries. As shown in Fig. 2(h), the measured Ga-content $x_{EDS}$ is nearly the same as the nominal values $x_{nominal}$ for all samples, and importantly the amount of Ga are pretty close between the single crystal and polycrystalline samples with the same $x$.

XPS spectra for both single crystal and polycrystalline samples with $x$=0, 0.05, and 0.09 are shown in Fig. 3. By fitting the data with multiple valance states of Ir ions, the presence of $Ir^{5+}$ in the doped samples is identified, and the content of $Ir^{5+}$ is increased with increasing $x$. Importantly, the content of $Ir^{5+}$ of the samples with the same $x$ are quite close, as listed in Table I, suggesting the same Ir valance state in the samples the same $x$. Regarding the tiny amount of $Ir^{5+}$ (~3%) in the non-doped samples, it may due to the experimental resolution, noting that it is much smaller the value of another two samples with $x$=0.05 and 0.09.

It is known that oxygen vacancies could affect the properties of $Sr_2IrO_4$ evidently [22,37]. Nevertheless, the influence of such non-stoichiometry can be neglected in the present work. First, $Sr_2IrO_4$ is expected to have a high $T_N$~240 K only if the sample has no (or tiny) oxygen vacancies [37]. In our work, both single crystal and polycrystalline $Sr_2IrO_4$ show antiferromagnetic transition at $T_N$~240 K (shown below). Second, it was found that resistivity can be reduced drastically, and a metallic state can be obtained by small amount of oxygen vacancies in $Sr_2IrO_4$ [22]. In the present work, both single crystal and polycrystalline $Sr_2IrO_4$ show insulating transport in the entire temperature range, and the magnitude of resistivity is similar to the literatures. Moreover, we annealed several single crystal samples with $x$=0, 0.05, and 0.09 at 700 °C for 72 hours in an oxygen atmosphere, and found that there is no apparent difference in electric transport and the antiferromagnetic transition temperature $T_N$ between the samples before and after annealing.

**B. Magnetic properties**

The different structure parameters may lead to different properties between the two groups of samples, since crystalline structure is important to the electronic and magnetic structure of $Sr_2IrO_4$. Because of the collective canting of $J_{eff}$=1/2 moments, there exists net magnetic moment $\mu_{net}$ in each $IrO_2$ layer. By applying $H$//$ab$-plane, the $J_{eff}$=1/2 moments and thereby $\mu_{net}$ can be reversed, resulting in an AFM to weak-ferromagnetic (wFM) transition (also usually called flop-transition) [38]. The wFM phase is typically featured by an abrupt $M$-enhancement at $T_N$ in $M(T)$ curves, which can be seen for all samples shown in Fig. 4. For the non-doped samples, including both single crystal and polycrystalline $Sr_2IrO_4$, the AFM transition is identified at $T$=$T_N$~240 K, in agreement with previous reports [2,21]. Upon Ga-doping, $T_N$ of the single crystals is moved down from $T_N$~240 K at $x$=0 to $T_N$~176 K at $x$=0.09, evidencing significant suppression of the AFM phase. However, for the polycrystalline samples, the AFM order behaves quite robust against Ga-doping, and there is no discernible decrease in $T_N$. For instance, at $x$=0.09, $T_N$ of the polycrystalline sample is about 64 K higher than the single crystal counterpart. The large difference in $T_N$ between the two groups of samples is summarized in Fig. 4(c).

The different magnetic properties between the two groups of samples is further evidenced by $M(H)$ measurements. Fig. 5(a) and (b) present $H$ dependence of $M_a$ ($H//a$-axis) and $M_c$ ($H//c$-axis) for the single crystals, respectively. As expected, Sr$_2$IrO$_4$ single crystal exhibits very thin $M_a(H)$ curve with clear magnetic saturation. The maximum magnetization $M_s$ is estimated to 0.093 $\mu_B$/Ir at $H$=5 T. Much smaller magnetization $M_c$ is identified, confirming the anisotropic magnetism of Sr$_2$IrO$_4$. Upon Ga-doping, both $M_{ab}$ and $M_c$ are reduced gradually, and a magnetic anisotropy $M_{ab}/M_c$~3 at $H$=5 T is derived for all samples, consistent with previous works [21,39].

Different $M(H)$ curves are observed for the group of polycrystalline samples, which all exhibit obvious hysteresis with large coercive field ($H_c$), shown in Fig. 5(c). For the non-doped sample, while the measured $M$ is as small as the single crystal counterpart, its $H_c$ is as huge as ~0.5 T, marking the significantly enhanced magnetic switching barrier probably due to grain boundaries. Similar phenomenon has been reported previously [33]. Increasing $x$ further enhances the coercive field to $H_c$~1 T at higher doping content, which is about two orders of magnitude larger than that of the single crystal samples. In spite of the large difference in $H_c$, $M$ of the polycrystalline samples decreases with $x$ continuously, resembling the case of single crystals as summarized in Fig. 5(c).

## C. Electric resistivity

Electric transport of the samples is also modulated drastically by Ga-doping. As shown in Fig. 6, both single crystal and polycrystalline Sr$_2$IrO$_4$ exhibit the same insulating transport, and the $\rho(T)$ curves take comparable enhancement of about five orders of magnitude when $T$ ranges from 300 K to 30 K. With Ga-doping, $\rho$ is reduced greatly for both groups of samples, which can be up to six orders of magnitude at low-$T$ region. As revealed by the XPS and Hall-effect measurements, Ga-doping is highly efficient in enhancing hole carrier density of the materials [30]. For instance, at $T$=300 K, the hole carrier density is increased considerably from $3.7 \times 10^{20}$ /cm$^3$ at $x$=0 to $1.6 \times 10^{21}$ /cm$^3$ at $x$=0.05, which would certainly promote the conduction significantly. For the polycrystalline samples, $\rho(T)$ curve is suppressed as $x\leq0.05$, and then enhanced quickly as $x$ is further increased to $x=0.15$. Clearly, the sample with $x=0.15$ has a $\rho(T)$ curve even higher than that of the non-doped $Sr_2IrO_4$. Fig. 6(c) presents $x$ dependence of $\rho$ taken at $T=50$ K. A striking rebound is seen in $\rho(x)$ as $x>0.05$ for the group of polycrystalline samples. This is in accordance with the large decrease in $\phi(x)$ shown in Fig. 1(g), reminiscent of the close correlation between the band-gap and the Ir-O-Ir bond angle $\phi$. For the single crystal samples, $\phi(x)$ just show slight decrease as $x>0.05$, and indeed $\rho(x)$ decreases modestly at relatively high doping content.

## IV. Discussion

The above experimental results have described strikingly different physical properties between the two groups of $Sr_2Ir_{1-x}Ga_xO_4$. In particular, the polycrystalline samples have much higher $T_N$ than the single crystals. In the following, we will mainly focus on discussing the structural origin of the discrepancies between the two groups of samples.

As aforementioned, the Ir-O-Ir bond angle $\phi$ has been the major lattice parameter that most frequently used to understand properties of $Sr_2IrO_4$. According to first principles calculations, the band gap of $Sr_2IrO_4$ can be widened (narrowed) as $\phi$ is reduced (increased) [9]. This effect will coexist and compete with the enhanced carrier density, and thus contributing to the transport of the materials. In the present work, it is identified that the reduction in $\rho$ is up to six orders of magnitude for both groups of samples. This is remarkable indeed, and is far larger than that of electrolyte gating experiments with $Sr_2IrO_4$ [40,41]. On the contrary, it is noted that electrolyte gating can inject far more carriers into $Sr_2IrO_4$, in comparison with Ga-doping [30]. Therefore, it is not proper to solely assign the remarkable reduction in $\rho$ to the enhanced carrier density. And, contribution arising from the increased $\phi$ at $x<0.07$ should be essentially taken into account, which would shrink the band gap, and thus active the conduction together with the enhanced hole carrier density. At $x>0.07$, $\phi$ is decreased, and then the band gap is expected to be expanded. This will strongly compete with the continuous enhancement of carrier density. As a consequence, $\rho$ will not decrease as striking as before, and may even show an increase if the reduction in $\phi$ is sufficiently large. This is in good agreement with the derived $\rho(x)$ data in Fig. 6(c), confirming the important role of $\phi$ in tuning electric transport of $Sr_2IrO_4$.

However, there is no discernible correlation between $T_N$ and $\phi$, since the two parameters evolve with $x$ in totally different ways. In fact, the in-plane Ir-O-Ir bond angle $\phi$ is more relevant to exchange interactions within the basal plane, while the long range AFM order is triggered by the interlayer coupling [19,20,24,42-44]. Because magnetic exchange interactions of $Sr_2IrO_4$ are sensitive to the distance of neighboring Ir ions, the $c/a$ ratio would be more relevant to the delicate balance of multiple exchange interactions in $Sr_2IrO_4$. For instance, with decreasing $c/a$, the neighboring layers are getting closer, and thus strengthening the interlayer coupling. This consequently tends to enhance $T_N$. In the present work, sizable reduction in $c/a$ is observed for the polycrystalline samples. This explains the commonly higher $T_N$ of the polycrystalline samples, in comparison with the single crystals. This scenario, illustrating intimate correlation between $T_N$ and the $c/a$ ratio, is solidly supported many previous studies. Fig. 7(a) summarizes the $c/a$ data of a series of samples with different dopants. Isoelectronic substitution of Ba and Ca for Sr provides a unique route to solely change the lattice of $Sr_2IrO_4$. It was reported that $(Sr,Ba,Ca)_2IrO_4$ samples just show slight variation in $c/a$ (less than 0.2%), and nearly unchanged $T_N$ [28,36]. With the application of pressure on $Sr_2IrO_4$, the $c/a$ ratio is found to increase evidently, opposite to the observation of present work. Indeed, $T_N$ is suppressed significantly by pressure [45,46]. This is in accordance with our proposed scenario. In Fig. 7(b), $\Delta T_N$ is plotted as a function of $\Delta(c/a)$, where $\Delta T_N$ and $\Delta(c/a)$ represent the difference in $T_N$ and $c/a$ between the polycrystalline and single crystal samples, respectively. Clearly, a higher $T_N$ can be achieved as the $c/a$ ratio is reduced in the $J_{eff}=1/2$ antiferromagnets.

At last, we qualitatively discuss the origin of the different lattice parameters between the two groups of samples. In $Sr_2IrO_4$, the rotation angle of IrO6 octahedra is determined by the in-plane bond-lengths of SrO and IrO2 layers; Ir-O bond length is too large to fit with the Sr-O layer and thus IrO6 need to rotate to compensate the difference. Ga doping induces hole-doping, and thus increases the Ir valence state, confirmed by the XPS measurements. Because of that, Ir-O bond length is reduced, and therefore the mismatch of SrO and IrO2 layers is partially reduced, leading to a smaller rotation angle $\varphi$. This is revealed by the continuous decrease in $l$ and $\varphi$ as $x<0.07$ shown in Fig. 1(f) and (h), respectively. However, further increasing $x$ causes clear upturn in $l(x)$ and $\varphi(x)$. This indicates a significant contribution from Ga-O bonds which tend to enhance the overall in-plane bond length, probably due to the smaller electrostatic attraction of $Ga^{3+}$ compared with

$Ir^{4+/5+}$. The competition of these factors is balanced by the Ga-content $x$, which determines the variation in $l(x)$ and $\varphi(x)$, shown in Fig. 1.

While the cooperation between $\phi$ and $l$ is responsible for the slight increase in the in-plane lattice constant $a$, the shrink of the $c$-axis suggests that the $IrO_2$ layers get closer upon Ga-doping. In particular, for the group of polycrystalline samples, the $c$-value is shortened as large as 1.6% at $x=0.15$, far larger than the single crystals. As a consequence, the cell volume of polycrystalline samples is reduced by ~1% at $x=0.15$, while it remains unchanged till $x=0.09$ for the single crystals. Such remarkable discrepancy implies that there exists additional compressive strain along the $c$-axis in the polycrystalline samples. The compressive strain should not be ascribed to differences of chemical composition between the two groups of samples, which can be excluded by the EDS and XPS characterizations. Alternatively, it may arise from the massive grain boundaries where grains with different orientations are merged in the polycrystalline samples. In a very recent experimental work [47], it revealed that the presence of strain at grain boundaries can modify the lattice parameters of $La_2CuO_4$ (isostructural to $Sr_2IrO_4$), in agreement with our observations.

To confirm the possible effect at grain boundaries, two single crystal samples with $x=0.05$ and 0.09 were crashed carefully, and the obtained powder was pressed into pellets and re-sintered at 1000 °C for 72 hours in air. The pellets were covered with the remained powder during the sintering. Preferred orientation still exists in the re-sintered samples, and therefore additional functions were used during the structural refinements. The derived lattice parameters are listed in Table II. In comparison with the pristine single crystals, the re-sintered polycrystalline samples commonly have smaller $c/a$ ratio, mainly due to compression of the $c$-axis. From the measured $M(T)$ data shown in Fig. 8, it is seen that the polycrystalline samples have obviously higher $T_N$ than the corresponding single crystals, i.e. $T_N$ is enhanced by ~10 K and ~30 K at $x=0.05$ and 0.09, respectively. These results indeed suggest that by introducing grain boundaries into the samples, the lattice parameters and antiferromagnetic transition $T_N$ can be modulated. Moreover, in the samples with higher Ga-content, competition among the structural ingredients would be more complicated as Ga-O bonds are more rigorously involved. Therefore, the lattice of samples with larger $x$ would behave more sensitive to the grain boundaries. As shown in Table R1, structural variation of the sample with $x=0.09$ is indeed larger than $x=0.05$.

## V. Conclusion

In summary, we performed a comparison study on structure, magnetism, and electric transport of a series of single crystal and polycrystalline $Sr_2Ir_{1-x}Ga_xO_4$. The experimental results revealed very different physical properties between the two groups of samples, which are strongly associated with the crystalline structure. In particular, the Néel temperature $T_N$ of the group of polycrystalline samples are obviously higher than that of the single crystals, which can even reach up to $\Delta T_N$ ~64 K at $x$=0.09. Our structure analysis suggested that the $c/a$ ratio is critical for the remarkable enhancement in $T_N$, and the Ir-O-Ir bond angle $\phi$ is more relevant to the electric transport of $Sr_2Ir_{1-x}Ga_xO_4$.

**Acknowledgements:** This work was supported by the National Nature Science Foundation of China (Grant Nos. 11774106, 51431006, 51721001, and 51571152), Hubei Province Natural Science Foundation of China (Grant No. 2020CFA083), and the National Key Research Projects of China (Grant No. 2016YFA0300101).

**References:**

[1] J. G. Rau, E. K.-H. Lee, and H.-Y. Kee, Annual Review of Condensed Matter Physics **7**, 195 (2016).

[2] G. Cao and P. Schlottmann, Rep. Prog. Phys. **81**, 042502 (2018).

[3] C. Lu and J. M. Liu, Adv. Mater. **32**, e1904508 (2020).

[4] G. Cao *et al.*, npj Quantum Materials **5**, 26 (2020).

[5] J. Son *et al.*, npj Quantum Materials **4**, 17 (2019).

[6] Z. Wang, D. Walkup, Y. Maximenko, W. Zhou, T. Hogan, Z. Wang, S. D. Wilson, and V. Madhavan, npj Quantum Materials **4**, 43 (2019).

[7] B. J. Kim *et al.*, Phys. Rev. Lett. **101**, 076402 (2008).

[8] G. Jackeli and G. Khaliullin, Phys. Rev. Lett. **102**, 017205 (2009).

[9] S. J. Moon, H. Jin, W. S. Choi, J. S. Lee, S. S. A. Seo, J. Yu, G. Cao, T. W. Noh, and Y. S. Lee, Phys. Rev. B **80**, 195110 (2009).

[10] B. Kim, P. Liu, and C. Franchini, Phys. Rev. B **95**, 115111 (2017).

[11] P. Liu, S. Khmelevskyi, B. Kim, M. Marsman, D. Li, X.-Q. Chen, D. D. Sarma, G. Kresse, and C. Franchini, Phys. Rev. B **92**, 054428 (2015).

[12] A. W. Lindquist and H.-Y. Kee, Phys. Rev. B **100**, 054512 (2019).

[13] H. Zhang, K. Haule, and D. Vanderbilt, Phys. Rev. Lett. **111**, 246402 (2013).

[14] S. Boseggia *et al.*, J Phys. Condens. Matter **25**, 422202 (2013).

[15] F. Ye, S. Chi, B. C. Chakoumakos, J. A. Fernandez-Baca, T. Qi, and G. Cao, Phys. Rev. B **87**, 140406 (2013).

[16] C. Lu, B. Gao, H. Wang, W. Wang, S. Yuan, S. Dong, and J.-M. Liu, Adv. Funct. Mater. **28**, 1706589 (2018).

[17] I. Fina *et al.*, Nat. Commun. **5**, 4671 (2014).

[18] H. Wang, C. Lu, J. Chen, Y. Liu, S. L. Yuan, S. W. Cheong, S. Dong, and J. M. Liu, Nat. Commun. **10**, 2280 (2019).

[19] J. Kim *et al.*, Phys. Rev. Lett. **108**, 177003 (2012).

[20] S. Fujiyama, H. Ohsumi, T. Komesu, J. Matsuno, B. J. Kim, M. Takata, T. Arima, and H. Takagi, Phys. Rev. Lett. **108**, 247212 (2012).

[21] M. Ge, T. F. Qi, O. B. Korneta, D. E. De Long, P. Schlottmann, W. P. Crummett, and G. Cao, Phys.

Rev. B **84**, 100402 (2011).

[22] O. B. Korneta, T. Qi, S. Chikara, S. Parkin, L. E. De Long, P. Schlottmann, and G. Cao, Phys. Rev. B **82**, 115117 (2010).

[23] T. Thio and A. Aharony, Phys. Rev. Lett. **73**, 894 (1994).

[24] A. Lupascu *et al.*, Phys. Rev. Lett. **112**, 147201 (2014).

[25] X. Chen *et al.*, Phys. Rev. B **92**, 075125 (2015).

[26] F. Ye, X. Wang, C. Hoffmann, J. Wang, S. Chi, M. Matsuda, B. C. Chakoumakos, J. A. Fernandez-Baca, and G. Cao, Phys. Rev. B **92**, 201112 (2015).

[27] S. J. Yuan, S. Aswartham, J. Terzic, H. Zheng, H. D. Zhao, P. Schlottmann, and G. Cao, Phys. Rev. B **92**, 245103 (2015).

[28] X. Chen and S. D. Wilson, Phys. Rev. B **94**, 195115 (2016).

[29] T. F. Qi, O. B. Korneta, L. Li, K. Butrouna, V. S. Cao, X. Wan, P. Schlottmann, R. K. Kaul, and G. Cao, Phys. Rev. B **86**, 125105 (2012).

[30] H. Wang, W. Wang, N. Hu, T. Duan, S. Yuan, S. Dong, C. Lu, and J.-M. Liu, Phys. Rev. Appl. **10**, 014025 (2018).

[31] K. Horigane *et al.*, Phys. Rev. B **97**, 064425 (2018).

[32] X. Chen *et al.*, Phys. Rev. B **92**, 075125 (2015).

[33] C. Cosio-Castaneda, G. Tavizon, A. Baeza, P. de la Mora, and R. Escudero, J Phys.: Condens. Matter **19**, 446210 (2007).

[34] M. K. Crawford, M. A. Subramanian, R. L. Harlow, J. A. Fernandez-Baca, Z. R. Wang, and D. C. Johnston, Phys. Rev. B **49**, 9198 (1994).

[35] M. K. Crawford, M. A. Subramanian, R. L. Harlow, J. A. Fernandez-Baca, Z. R. Wang, and D. C. Johnston, Phys. Rev. B **49**, 9198 (1994).

[36] H. D. Zhao, J. Terzic, H. Zheng, Y. F. Ni, Y. Zhang, F. Ye, P. Schlottmann, and G. Cao, J Phys. Condens. Matter **30**, 245801 (2018).

[37] N. H. Sung, H. Gretarsson, D. Proepper, J. Porras, M. Le Tacon, A. V. Boris, B. Keimer, and B. J. Kim, Philosophical Magazine **96**, 413 (2016).

[38] B. J. Kim, H. Ohsumi, T. Komesu, S. Sakai, T. Morita, H. Takagi, and T. Arima, Science **323**, 1329 (2009).

[39] J. C. Wang *et al.*, Phys. Rev. B **92**, 214411 (2015).

[40] C. Lu, S. Dong, A. Quindeau, D. Preziosi, N. Hu, and M. Alexe, Phys. Rev. B **91**, 104401 (2015).

[41] J. Ravichandran, C. R. Serrao, D. K. Efetov, D. Yi, Y. S. Oh, S. W. Cheong, R. Ramesh, and P. Kim, J Phys. Condens. Matter **28**, 505304 (2016).

[42] L. Hao *et al.*, Phys. Rev. Lett. **119**, 027204 (2017).

[43] V. M. Katukuri, V. Yushankhai, L. Siurakshina, J. van den Brink, L. Hozoi, and I. Rousochatzakis, Phys. Rev. X **4** (2014).

[44] T. Takayama, A. Matsumoto, G. Jackeli, and H. Takagi, Phys. Rev. B **94**, 224420 (2016).

[45] C. Chen *et al.*, Phys. Rev. B **101**, 144102 (2020).

[46] D. Haskel *et al.*, Phys. Rev. Lett. **124**, 067201 (2020).

[47] J. Wang, C. Chen, B. Huang, J. Cao, L. Li, Q. Shao, L. Zhang, and X. Huang, Nano Lett. 21, 980 (2021).

**Figure captions:**

Figure 1. (a) Rietveld refinement of $Sr_2IrO_4$ single crystal. Evaluated lattice parameters: (b) $a$, (c) $c$, (d) variation in cell volume $\Delta V$, (e) the $c/a$ ratio, (e) the Ir-O2-Ir bond length $l$, (f) the Ir-O2-Ir bond angle $\phi$, and (g) rotation angle $\varphi$ of oxygen octahedra for the polycrystalline (blue squares) and single crystal (red dots) $Sr_2Ir_{1-x}Ga_xO_4$.

Figure 2. (a) and (b) show SEM images of the single crystal and polycrystalline $Sr_2Ir_{0.93}Ga_{0.07}O_4$, respectively. The inset in (b) shows a magnified image for the polycrystalline sample with $x=0.07$, and the corresponding elemental mappings of Sr, Ir, Ga and O are shown in (c)-(f), respectively. A EDS spectra of sample with $x=0.07$ is show in (g). (h) Measured Ga content $x_{EDS}$ as a function of nominal value $x_{nominal}$.

Figure 3. Measured XPS spectra for samples with $x=0$, 0.05, and 0.09. The spectra are fitted with multiple valance states of Ir.

Figure 4. Temperature dependence of magnetization for (a) single crystal $Sr_2Ir_{1-x}Ga_xO_4$ ($0 \leq x \leq 0.09$), and (b) for polycrystalline samples with $0 \leq x \leq 0.15$. (c) The Néel temperature $T_N$ as a function of Ga-content $x$ for both groups of samples.

Figure 5. Measured magnetization as a function of $H$ at $T=10$ K. (a) In-plane magnetization $M_a(H)$ and (b) out-of-plane magnetization $M_c(H)$ of the single crystals. (c) $M(H)$ curves for the polycrystalline samples. (d) $x$ dependence of measured magnetization $M_s$ at $H=5$ T.

Figure 6. Temperature dependence of $\rho$ for (a) the single crystals, and (b) the polycrystalline samples. (c) Estimated $\rho$ as a function of $x$ at $T=50$ K for both groups of samples.

Figure 7. (a) The $c/a$ ratio as a function of doping content $x$ for several groups of samples with different dopants. The data of Ca and Ba doped $Sr_2IrO_4$ were taken from Ref[28] and [36]. (b) $\Delta T_N$ as a function of the $\Delta(c/a)$ ratio, $\Delta T_N$ and $\Delta(c/a)$ represent the difference in $T_N$ and $c/a$ between the polycrystalline and single crystal samples, respectively.

Figure 8. Measured in-plane magnetization $M_a$ as a function of $T$ for the samples with $x$=0.05 and 0.09. The re-sintered polycrystals were obtained from corresponding single crystals.

Table I. XPS fitting results for samples with $x$=0, 0.05, and 0.09.

| $Sr_2Ir_{1-x}Ga_xO_4$ | $x=0$ | | $x=0.05$ | | $x=0.09$ | |
|---|---|---|---|---|---|---|
| | $Ir^{4+}$ | $Ir^{5+}$ | $Ir^{4+}$ | $Ir^{5+}$ | $Ir^{4+}$ | $Ir^{5+}$ |
| Single crystal | 97.16% | 2.84% | 84.96% | 15.04% | 82.21% | 17.79% |
| Polycrystal | 96.82% | 3.18% | 85.34% | 14.66% | 83.65% | 17.35% |

Table II. Obtained lattice parameters from the structure refinements of re-sintered polycrystalline samples.

| | | $a$ (Å) | $c$ (Å) | $c/a$ | $\phi$ (°) |
|---|---|---|---|---|---|
| $x=0.05$ | Single crystal | 5.4970 | 25.752 | 4.685 | 159.8 |
| | Polycrystal | 5.4996 | 25.716 | 4.676 | 160.8 |
| $x=0.09$ | Single crystal | 5.4993 | 25.740 | 4.681 | 158.2 |
| | Polycrystal | 5.5002 | 25.681 | 4.669 | 161.5 |

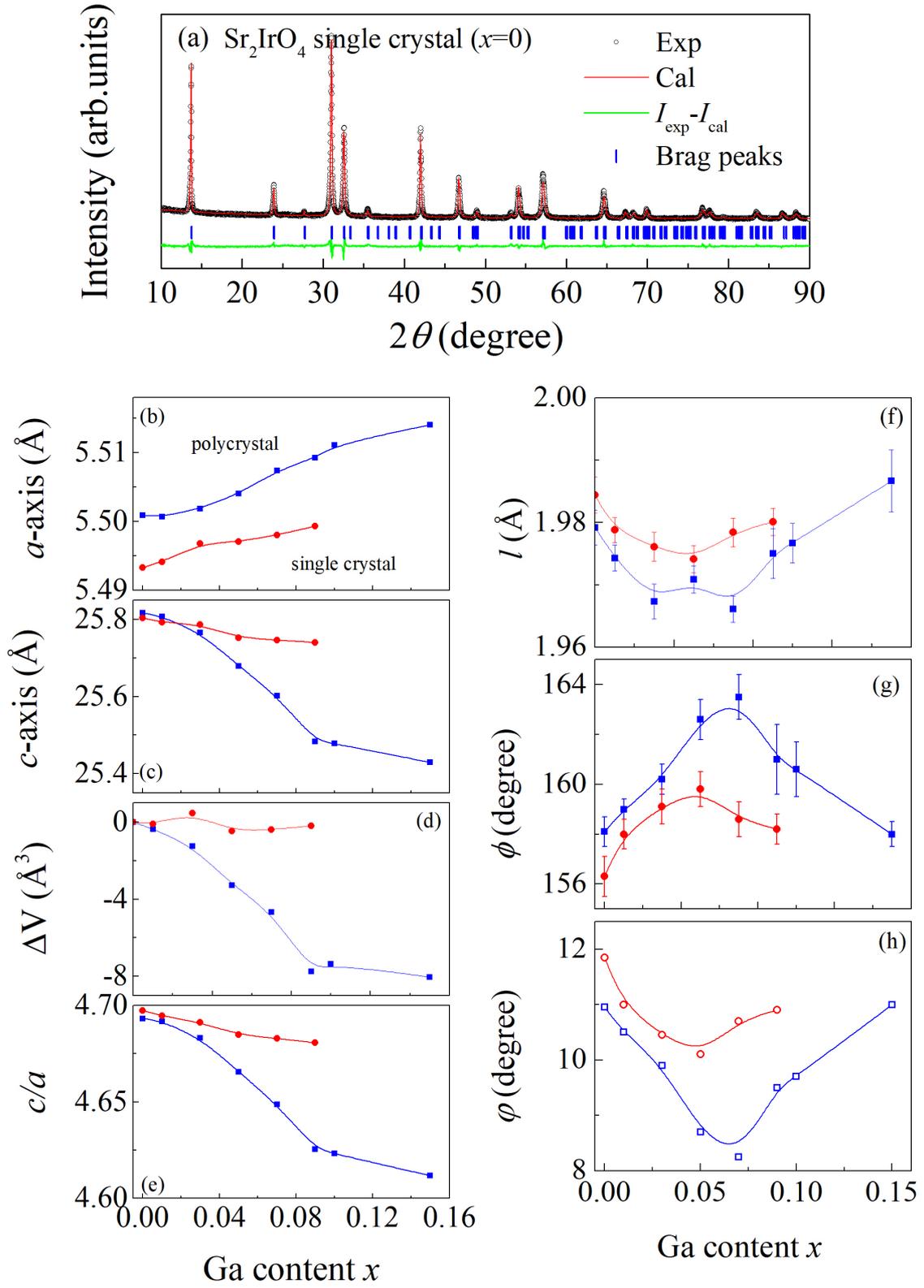

Figure 1

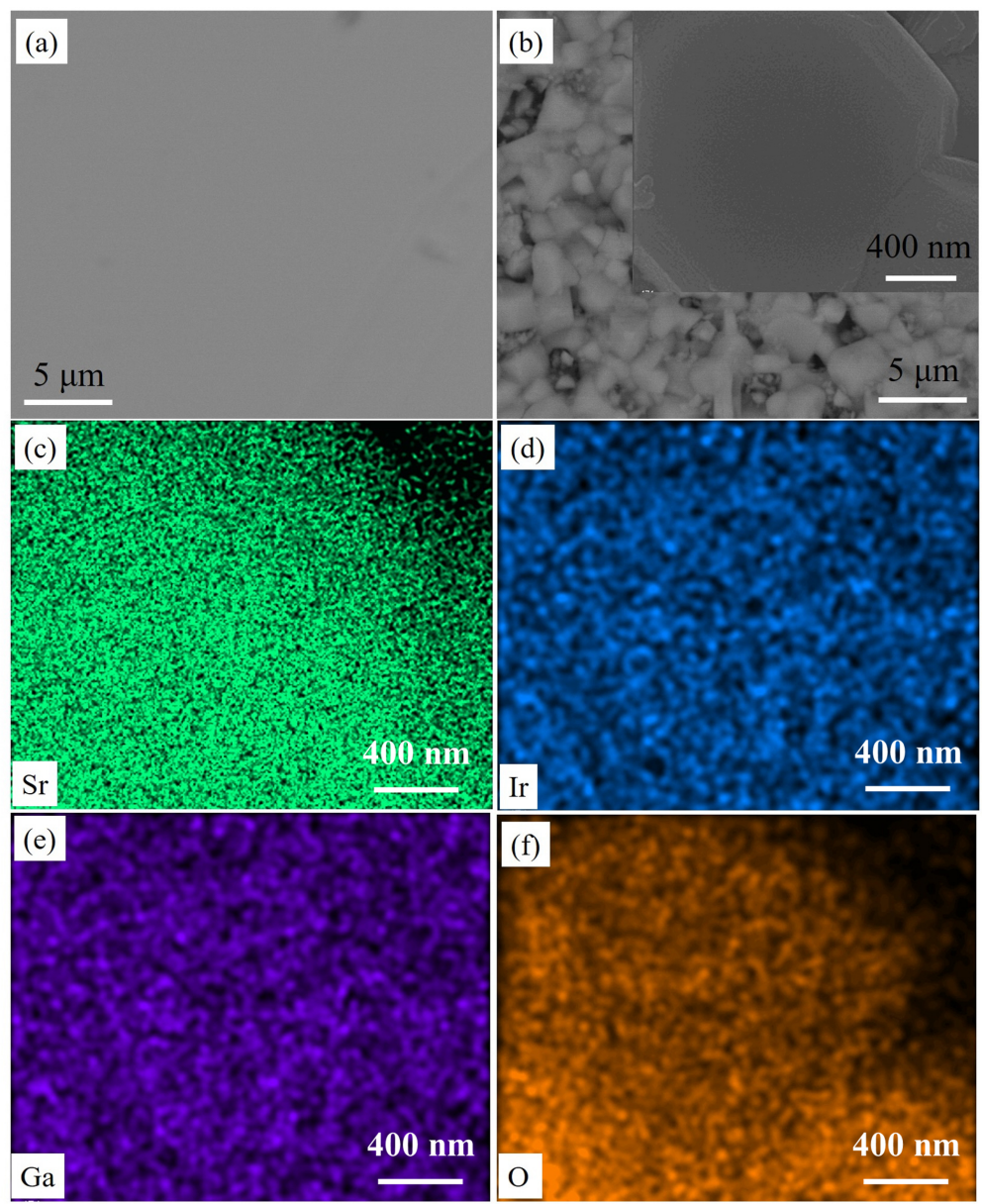

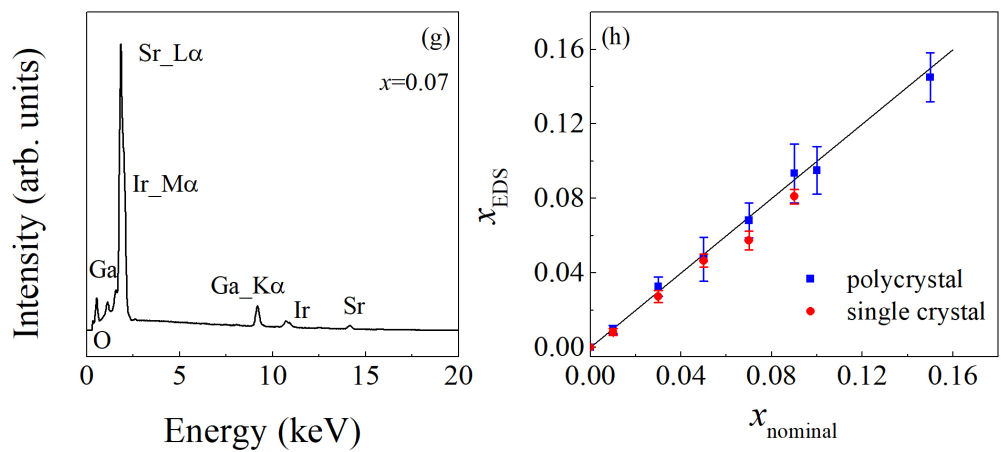

Figure 2

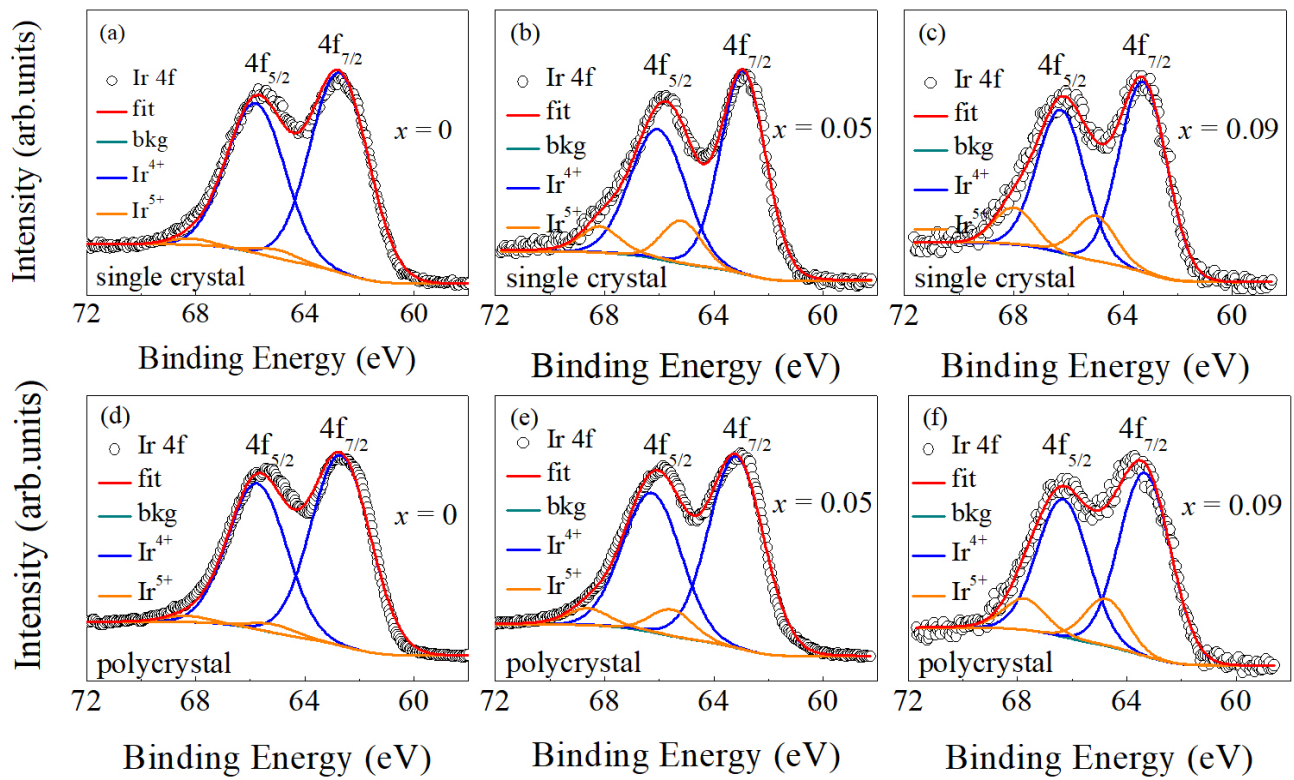

Figure 3

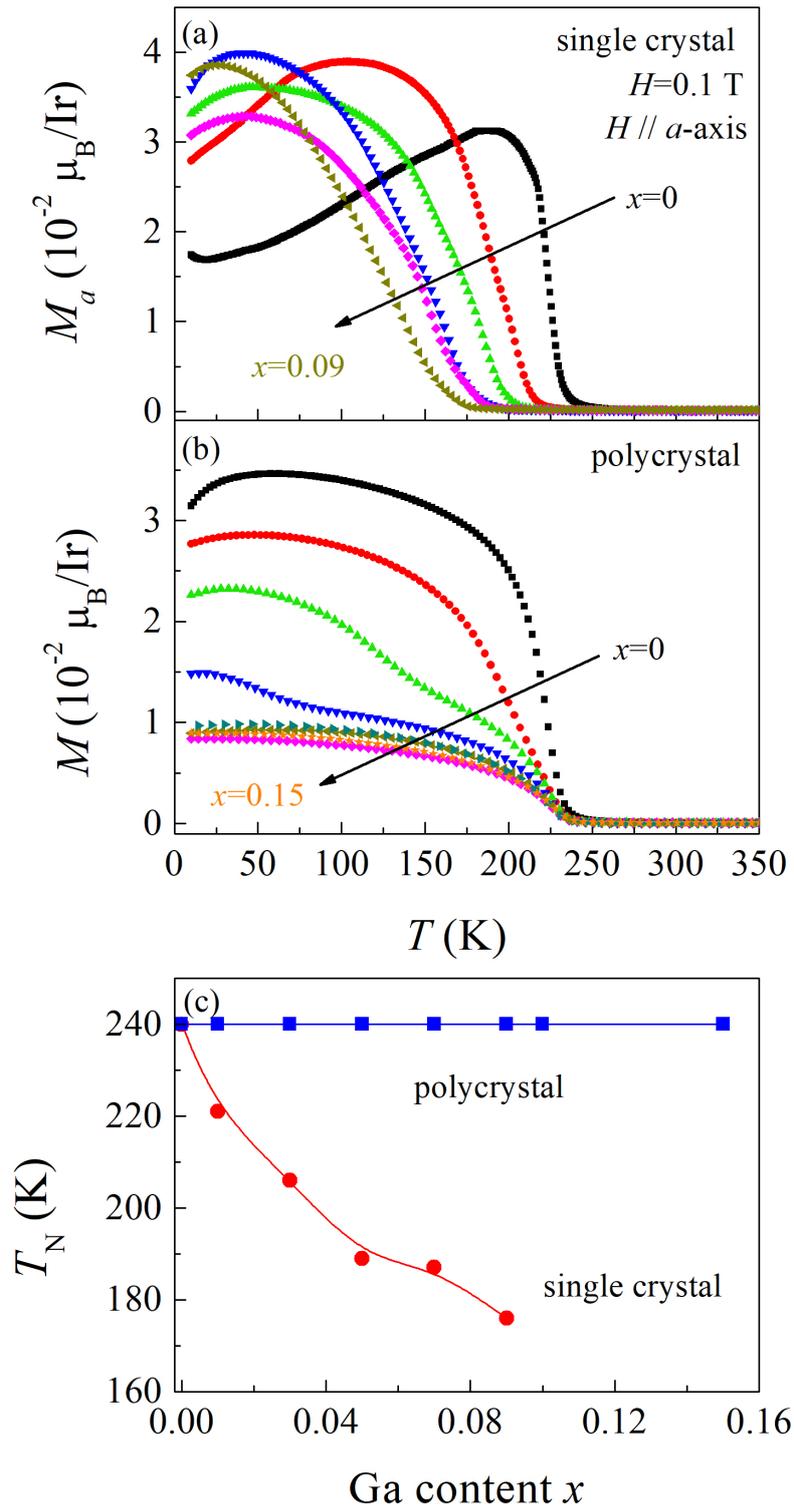

Figure 4

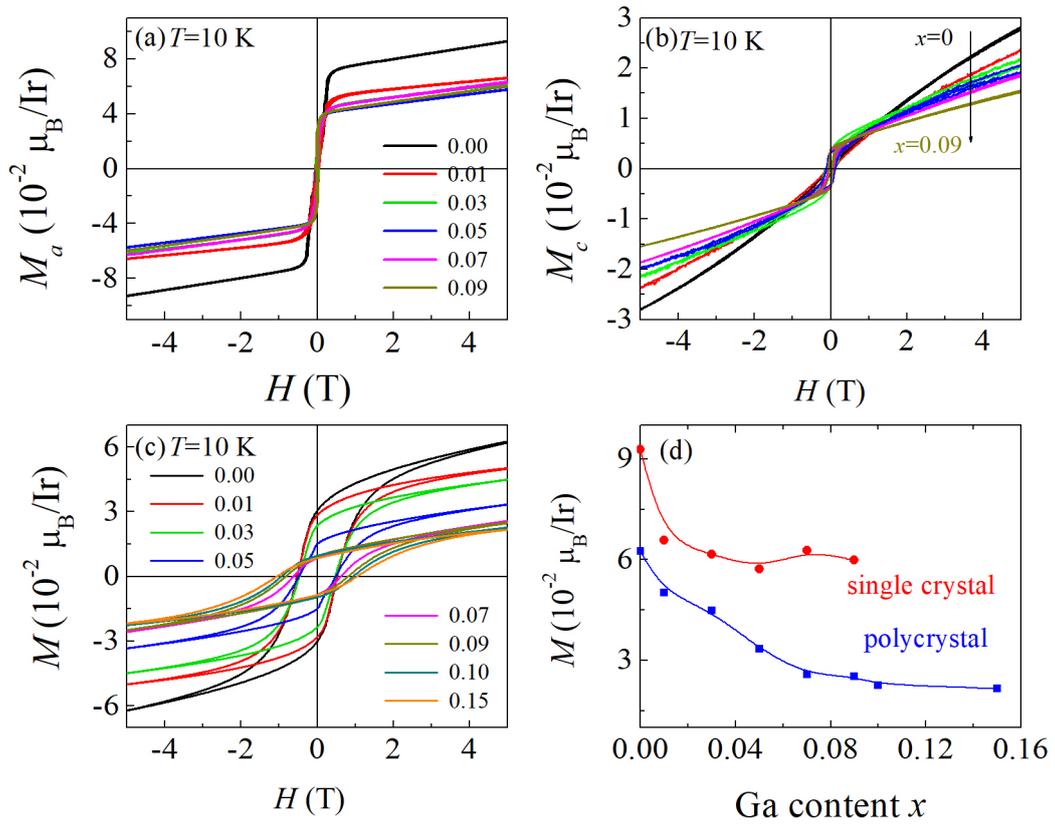

Figure 5

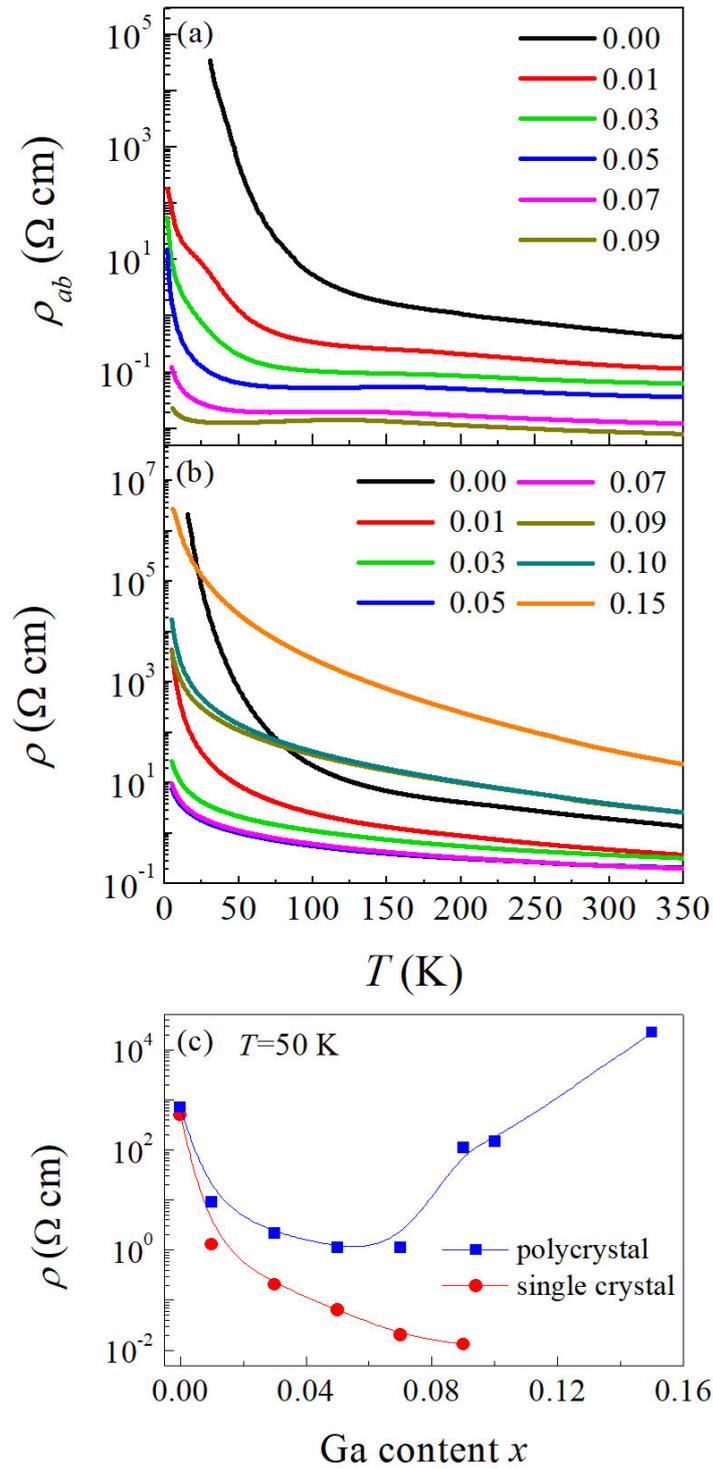

Figure 6

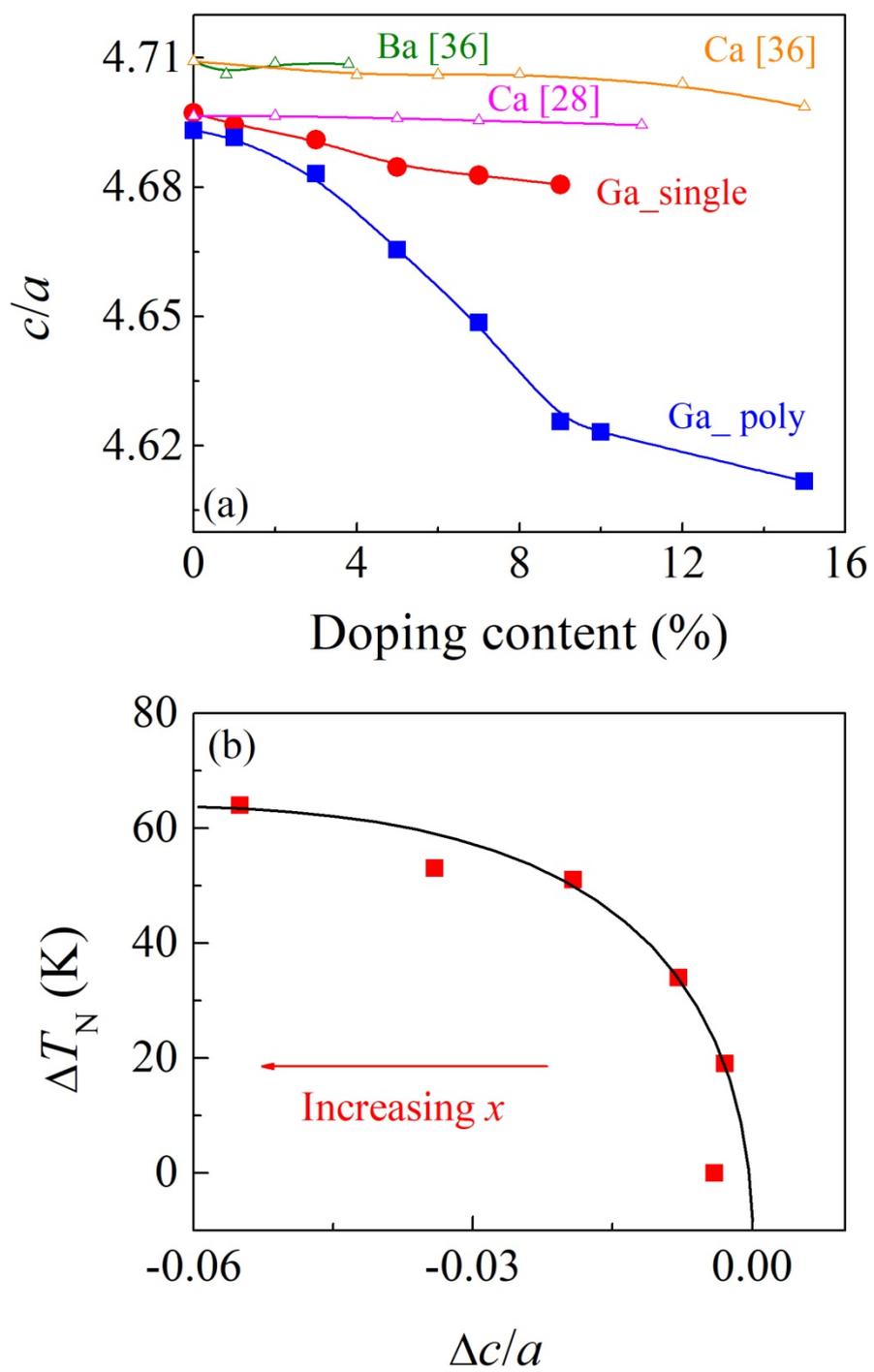

Figure 7

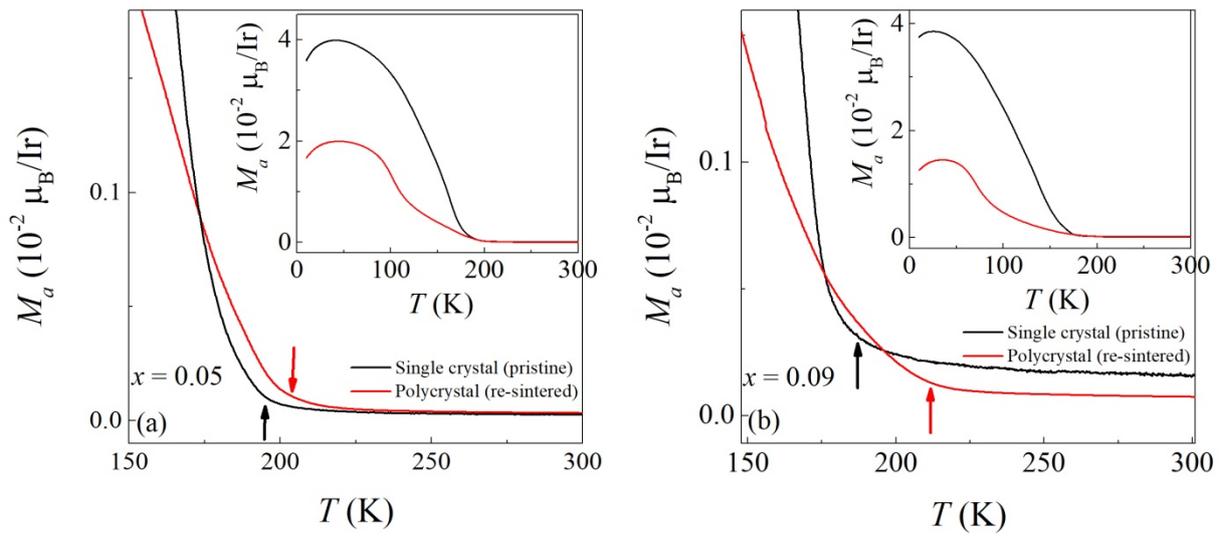

Figure 8